\documentclass[journal,review,submit,pdftex,oneauthors]{mdpi} 
\firstpage{1} 
\pubvolume{1}
\issuenum{1}
\articlenumber{0}
\pubyear{2025}
\copyrightyear{2025}
\datereceived{ } 
\daterevised{ } % Comment out if no revised date
\dateaccepted{ } 
\datepublished{ } 
\hreflink{https://doi.org/} % If needed use \linebreak
\Title{Stellar-Mass Black Holes}
\TitleCitation{Stellar-Mass Black Holes}

\Author{Cosimo Bambi $^{1,2}$\orcidA{}}

\AuthorNames{Cosimo Bambi}

\isAPAStyle{%
       \AuthorCitation{Bambi, C.}
         }{%
        \isChicagoStyle{%
        \AuthorCitation{Bambi, Cosimo.}
        }{
        \AuthorCitation{Bambi, C.}
        }
}

\address{%
$^{1}$ \quad Center for Astronomy and Astrophysics, Center for Field Theory and Particle Physics, and Department of Physics, Fudan University, Shanghai 200438, China; bambi@fudan.edu.cn\\
$^{2}$ \quad School of Natural Sciences and Humanities, New Uzbekistan University, Tashkent 100007, Uzbekistan}

\abstract{Stellar-mass black holes ($3$~$M_\odot \lesssim M_{\rm BH} \lesssim 150$~$M_\odot$) are the natural product of the evolution of heavy stars ($M_{\rm star} \gtrsim 20$~$M_\odot$). In our Galaxy, we expect $10^8$-$10^9$~stellar-mass black holes formed from the gravitational collapse of heavy stars, but currently we know fewer than 100~objects. We also know $\sim 100$~stellar-mass black holes in other galaxies, most of them discovered by gravitational wave observatories in the past 10~years. The detection of black holes is indeed extremely challenging and possible only in very special cases. This article is a short review on the physics and astrophysics of stellar-mass black holes, including Galactic and extragalactic black holes in X-ray binaries, black holes in astrometric binaries, isolated black holes, and black holes in compact binaries. The article also addresses some important open issues and introduces the idea of a possible interstellar mission to the closest black hole.}

\keyword{stellar-mass black holes; black hole binaries; binary black holes; isolated black holes.}

\begin{document}

%%%%%%%%%%%%%%%%%%%%%%%%%%%%%%%%%%%%%%%%%%

\section{Introduction}

A {\it black hole} is a region of the spacetime in which gravity is so strong that nothing can escape to the exterior region; the {\it event horizon} is the boundary of the black hole. In other words, the event horizon acts as a one-way membrane: massive and massless particles can cross the event horizon from the exterior region to the black hole but nothing can cross the event horizon from the black hole to the exterior region. For a more rigorous definition, see, for instance, Refs.~\cite{Misner:1973prb,Bambi:2017khi}.

In General Relativity, black holes are relatively simple objects and are completely characterized by a small number of parameters. This is the celebrated result of the {\it no-hair theorem}, which is actually a family of theorems and they hold under specific assumptions (number of the spacetime dimensions, absence of spacetime singularities and closed time-like curves on and outside of the event horizon, etc.)~\cite{Carter:1971zc,Robinson:1975bv,Chrusciel:2012jk}. The simplest black hole solution is the {\it Schwarzschild} spacetime, which describes a non-rotating and electrically uncharged black hole and is completely characterized by one parameter, the black hole mass ($M_{\rm BH}$). A non-rotating black hole with a non-vanishing electric charge is described by the {\it Reissner-Nordstr\"om} solution and has two parameters: the black hole mass ($M_{\rm BH}$) and the black hole electric charge ($Q_{\rm BH}$). A rotating black hole with a vanishing electric charge is instead described by the {\it Kerr} solution and has two parameters: the black hole mass ($M_{\rm BH}$) and the black hole spin angular momentum ($J_{\rm BH}$). Last, the general case of a rotating black hole with a non-vanishing electric charge is described by the {\it Kerr-Newman} solution, which has three parameters: the black hole mass ($M_{\rm BH}$), the black hole electric charge ($Q_{\rm BH}$), and the black hole spin angular momentum ($J_{\rm BH}$).

From a theoretical point of view, in General Relativity the black hole mass can assume any value, in the sense that there are no lower or upper bounds on the value of $M_{\rm BH}$ for the existence of a black hole. On the contrary, the electric charge and the spin angular momentum cannot be arbitrary and must satisfy the following constraint (in natural units in which $G_{\rm N} = c = 1$)
\begin{eqnarray}\label{eq-horizon}
\frac{Q_{\rm BH}^2}{2} + \sqrt{ \frac{Q_{\rm BH}^4}{4} + J_{\rm BH}^2 } \le M_{\rm BH}^2 \, .
\end{eqnarray}
Eq.~(\ref{eq-horizon}) is the condition for the existence of the event horizon: if Eq.~(\ref{eq-horizon}) is not satisfied, the Kerr-Newman solution does not describe a black hole but a naked singularity, which is normally thought to be impossible to create in Nature~\cite{Penrose:1969pc}.

The black holes that can be found in Nature must form through some physical process, which can further limit the ranges of $M_{\rm BH}$, $Q_{\rm BH}$, and $J_{\rm BH}$. Observations have discovered at least two classes of black holes:
\begin{enumerate}
\item {\it Stellar-mass black holes} have masses in the range $\sim 3$~$M_\odot$ to $\sim 150$~$M_\odot$. They are the natural product of the evolution of heavy stars. They are the topic of this review article and they will be discussed in the next sections.
\item {\it Supermassive black holes} have masses in the range $10^5$-$10^{10}$~$M_\odot$ and are found at the center of every middle-size and large galaxy ($M_{\rm galaxy} > 10^{11}$~$M_\odot$)\footnote{Exceptions may be possible: the galaxy A2261-BCG is one of the largest galaxies known but does not seem to host any supermassive black hole at its center~\cite{Postman:2012qq}.}~\cite{Kormendy:1995er}. In the case of small galaxies ($M_{\rm galaxy} < 10^{10}$~$M_\odot$), the situation is more controversial: some small galaxies host a supermassive black hole at their center while other small galaxies do not seem to have any supermassive black hole~\cite{Ferrarese:2006fd,Gallo:2007xq}. While heavy objects can naturally migrate to the center of a multi-body system, which explains why supermassive black holes are at the center of their host galaxies, we do not know exactly how they formed and evolved: for example, they could have formed as stellar-mass black holes from the collapse of heavy stars and grown later or they could have formed from the collapse of heavy clouds and been heavier than stellar-mass black holes from the very beginning. 
\end{enumerate}  
There is likely a third class of objects, the so-called {\it intermediate-mass black hole candidates}, with masses filling the gap between the stellar-mass and the supermassive black holes~\cite{Greene:2019vlv}. These black holes seem to be definitively more rare than those in the other two classes and currently there are no robust measurements of their masses, so it is possible that some of these objects are really intermediate-mass black holes but other objects are not. These three black hole classes (stellar-mass, supermassive, and intermediate-mass black holes) are sometimes referred to as {\it astrophysical} black holes, as they are thought to form by astrophysical processes\footnote{There are even theoretical models that predict that the supermassive black holes in galactic nuclei were produced in the early Universe and should thus be considered as primordial black holes~\cite{Dolgov:1992pu}.}.

{\it Primordial black holes} are an heterogeneous (an so far completely hypothetical) fourth class of black holes predicted by a number of different theoretical models~\cite{Byrnes:2025tji}. They are named ``primordial'' because they would be produced in the early Universe, before the first stars. They can be produced by a number of different mechanisms, with different mass ranges and distributions. Primordial black holes are sometimes referred to as {\it cosmological} black holes, as they can only form in the early Universe. So far there is no evidence of their existence, but recently they have attracted the interest of a large community.

The spacetime geometry around astrophysical black holes should be normally approximated well by the Kerr solution describing a stationary and uncharged black hole in vacuum. As soon as a black hole is formed, deviations from the Kerr metric can be quickly radiated away by the emission of gravitational waves~\cite{Price:1971fb}. Astrophysical objects can have a non-vanishing electric charge as a result of the large difference between the proton and electron masses, but for macroscopic black holes the equilibrium electric charge is completely negligible for the spacetime geometry~\cite{Bambi:2017khi,Bambi:2008hp}. The presence of accretion disks and/or nearby stars is also completely negligible near a black hole~\cite{Barausse:2014tra,Bambi:2014koa}: the masses of accretion disks are normally many orders of magnitude smaller than those of their black holes and disks are extended and have low densities; the gravitational field of nearby stars is also too weak to alter appreciably the Kerr geometry around a black hole.

In the case of the Kerr metric, Eq.~(\ref{eq-horizon}) reduces to $| a_* | \le 1$ ({\it Kerr bound}), where $a_*$ is the dimensionless spin parameter of the black hole defined as $a_* = J_{\rm BH} / M_{\rm BH}^2$. In Boyer-Lindquist coordinates, the radial coordinate of the event horizon of a Kerr black hole is
\begin{eqnarray}\label{eq-horizon2}
r_{\rm H} = r_{\rm g} \left( 1 + \sqrt{1 - a_*^2} \right) \, ,
\end{eqnarray}
where $r_{\rm g}$ is the gravitational radius of the black hole
\begin{eqnarray}\label{eq-grad}
r_{\rm g} = M_{\rm BH} = 14.8 \left( \frac{M}{10~M_\odot} \right)~{\rm km} \, .
\end{eqnarray}

The study of equatorial circular orbits in the Kerr spacetime has important applications in astrophysics~\cite{Bambi:2017khi}. In Newtonian gravity, circular orbits in the gravitational field of a point-like massive object are always stable. In the Kerr spacetime, this is not true: the innermost stable circular orbit (or ISCO) is the circular orbit on the equatorial plane separating stable orbits at larger radii from unstable orbits at smaller radii. In Boyer-Lindquist coordinates, the radial coordinate of the ISCO can be written in the following compact analytic form
\begin{eqnarray}\label{eq-isco}
r_{\rm ISCO} = r_{\rm g} \left[ 3 + Z_2 \mp \sqrt{\left( 3 - Z_1 \right)\left( 3 + Z_1 + 2 \, Z_2 \right)} \right] \, ,
\end{eqnarray}
where the sign $-$ ($+$) refers to co-rotating (counter-rotating) orbits and $Z_1$ and $Z_2$ are defined as
\begin{eqnarray}
Z_1 &=& 1 + \left( 1 - a_*^2 \right)^{1/3} 
\left[ \left( 1 + a_* \right)^{1/3} + \left( 1 - a_* \right)^{1/3} \right] \, , \\
Z_2 &=& \sqrt{3 \, a_*^2 + Z_1^2} \, .
\end{eqnarray}
For $a_* = -1$, 0, and 1, we have $r_{\rm ISCO} = 9 \, r_{\rm g}$, $6 \, r_{\rm g}$, and $r_{\rm g}$, respectively. 

\begin{figure}
\centering
\includegraphics[width=1.0\textwidth,trim=1.8cm 2.0cm 6.8cm 17.5cm,clip]{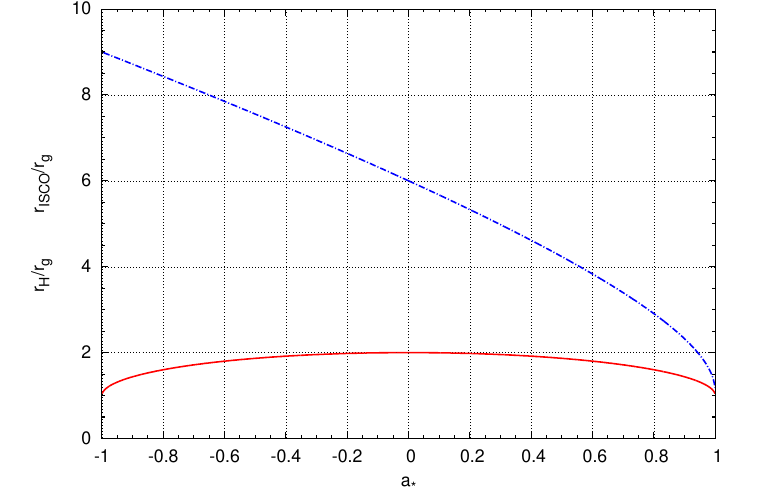}
\caption{Radial coordinate of the event horizon $r_{\rm H}$ (red solid curve) and of the ISCO radius $r_{\rm ISCO}$ (blue dashed-dotted curve) in the Kerr spacetime in Boyer-Lindquist coordinates as a function of the black hole spin parameter $a_*$. $a_* > 0$ corresponds to the case of co-rotating orbits, namely orbits with angular momentum parallel to the black hole spin. $a_* < 0$ is for counter-rotating orbits, namely orbits with angular momentum anti-parallel to the black hole spin.}
\label{f-isco}
\end{figure}

Fig.~\ref{f-isco} shows $r_{\rm H}$ and $r_{\rm ISCO}$ as a function of the black hole spin parameter $a_*$. The ISCO radius depends on the black hole spin and the angular momentum of the orbiting particle: $a_* > 0$ corresponds to the case of co-rotating orbits (orbits with angular momentum parallel to the black hole spin) and $a_* < 0$ corresponds to the case of counter-rotating orbits (orbits with angular momentum anti-parallel to the black hole spin).

%%%%%%%%%%%%%%%%%%%%%%%%%%%%%%%%%%%%%%%%%%

\section{Stellar-Mass Black Holes}

Stellar-mass black holes can be expected as the natural product of the evolution of heavy stars~\cite{Burrows:2023nlq,Burrows:2024wqv} and their existence is confirmed by observations. When a star exhausts all its nuclear fuel, it cools down, the thermal pressure of its particles cannot compensate any longer the star's own weight, and the body collapses. This is normally a violent process, which naturally ejects a significant fraction of material to space. 

As a star collapses, its density increases, and at some point electrons become degenerate: if the quantum pressure of electrons can stop the collapse, we have the formation of a white dwarf (this is the case of stars with $M_{\rm star} \lesssim 8$~$M_\odot$). The maximum mass for a white dwarfs is $\sim 1.4$~$M_\odot$ ({\it Chandrasekhar limit})~\cite{Chandrasekhar:1931ih}. If the collapsing part of the star exceeds the Chandrasekhar limit, the quantum pressure of electrons cannot stop the collapse, the density increases further, and it is energetically convenient to convert protons and electrons into neutrons. At sufficiently high densities, neutrons become degenerate: if the quantum pressure of neutrons can stop the collapse, we have the formation of a neutron star (this is the case of stars with $8$~$M_\odot \lesssim M_{\rm star} \lesssim 20$~$M_\odot$). The maximum mass for a neuron star is $\sim 3$~$M_\odot$ ({\it Oppenheimer-Volkoff limit})~\cite{Oppenheimer:1939ne,Rhoades:1974fn,Kalogera:1996ci,Lattimer:2012nd}. If the neutron star exceeds the Oppenheimer-Volkoff limit, there is no known mechanism capable of stopping the collapse and we have the formation of a black hole (the case of stars with $M_{\rm star} \gtrsim 20$~$M_\odot$). The value of the minimum mass of a stellar-mass black hole formed from the collapse of a star corresponds thus to the maximum mass of a neutron star.

The value of the maximum mass of a stellar-mass black hole formed from the collapse of a star depends on the mass of the progenitor star and its metallicity~\cite{Heger:2002by,Spera:2015vkd,Mapelli:2021taw}. Since heavy elements have larger cross sections than light elements, a higher metallicity increases the mass loss during the explosion of the star. Heavy stars with Solar metallicity can produce black holes with masses up to $\sim 20$~$M_\odot$ and the rest of the mass is ejected to space. Heavy stars with very low metallicity can collapse and produce black hole remnants with roughly half of their masses. The formation of black holes with masses between $\sim 60$ and $\sim 120$~$M_\odot$ (from metal-poor stars with masses between $\sim 120$ and $\sim 250$~$M_\odot$) is suppressed by pair-instability~\cite{Heger:2001cd,Belczynski:2016jno,Woosley:2016hmi,Woosley:2019nnp}, but there are uncertainties on this mass gap~\cite{Farmer:2019jed,Mapelli:2019ipt,Costa:2020xbc,Vink:2020nak}. Black holes with masses above $\sim 150$~$M_\odot$ are unlikely simply because stars with $M_{\rm star} \gtrsim 300$~$M_\odot$ would exceed their Eddington luminosity and therefore cannot form\footnote{Black holes with masses exceeding $\sim 150$~$M_\odot$ can still form from the merger of two black holes~\cite{LIGOScientific:2025rsn}.}. 

From population evolution studies, we expect that in our Galaxy there are around $10^8$-$10^9$ stellar-mass black holes formed from the gravitational collapse of heavy stars. For example, assuming that the maximum mass for neutron stars is 1.7~$M_\odot$, Timmes~et~al.~(1996)~\cite{Timmes:1995kp} predicted $1.4 \cdot 10^9$~stellar-mass black holes in the Galaxy. Today we know that the maximum mass for a neutron star is somewhat higher~\cite{Antoniadis:2013pzd}, but from the model in Timmes~et~al.~(1996) we can still expect around $1.0 \cdot 10^9$~stellar-mass black holes in the Galaxy. On the other hand, Olejak~et~al.~(2020)~\cite{Olejak:2019pln} predicted around $1.1 \cdot 10^8$~stellar-mass black holes in the Galaxy: they estimated that in the Galactic disk there are about $1.0 \cdot 10^8$~isolated black holes and about $8 \cdot 10^6$~black holes in binary systems, in the Galactic bulge there are about $1.7 \cdot 10^7$~isolated black holes and about $1 \cdot 10^6$~black holes in binary systems, and in the Galactic halo there are about $4 \cdot 10^6$~isolated black holes and about $5 \cdot 10^5$~black holes in binary systems\footnote{Most stellar-mass black holes are expected to be isolated rather than in a binary system with companion stars for the combination of three effects: $i)$ about 30\% of massive stars are isolated stars, without companions; $ii)$ in a close binary, the two stars may merge before the supernova explosion of one of them, and $iii)$ in a wide binary, the supernova explosion can easily break up the system, producing an isolated black hole.}

Despite we expect $10^8$-$10^9$ stellar-mass black holes in our Galaxy, as of now we know less than 100~objects. The {\it known} stellar-mass black holes in our Galaxy can be grouped as follows and will be discussed in the next subsections\footnote{We also know a number of stellar-mass black hole candidates, for which there is no unanimous consensus they are really black holes; see, for instance, Refs.~\cite{NGC3201a,NGC3201b,Thompson:2018ycv,OB110462,Mahy:2022tde,Gaia18ajz}.}:
\begin{enumerate}
\item $\sim 70$~black holes in X-ray binaries. 
\item 4~black holes in astrometric binaries.
\item 1~isolated black hole.
\end{enumerate}
We also know around 10 stellar-mass black holes in X-ray binaries and astrometric binaries in nearby galaxies (around 20 black holes if we include even weak candidates) and about 100 gravitational wave events associated to the coalescence of two stellar-mass black holes or the coalescence of a stellar-mass black hole and a neutron star. These objects will also be discussed briefly in the next subsections. 

\begin{figure}[t]
\centering
\includegraphics[width=1.0\textwidth]{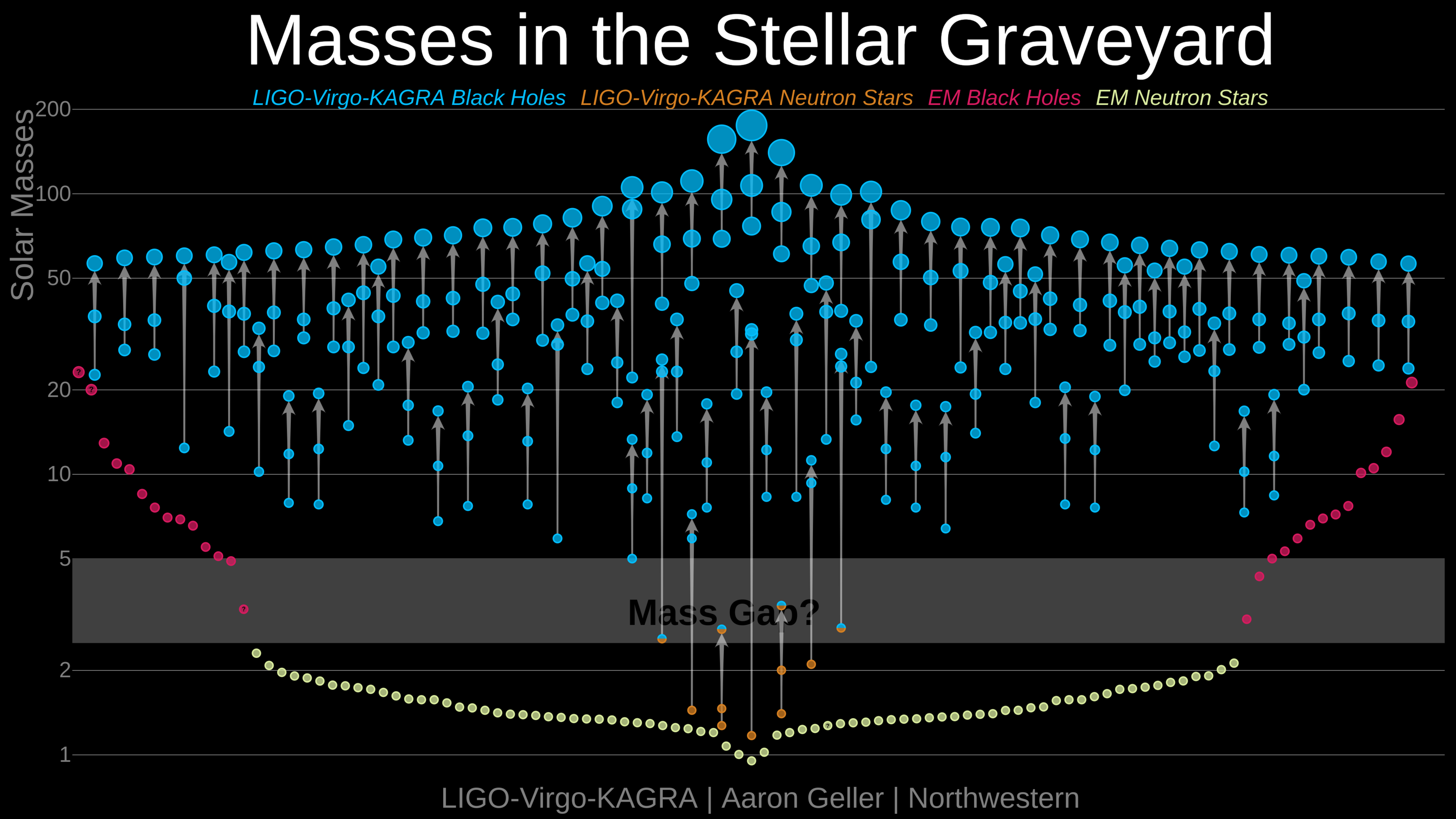}
\caption{Stellar-mass black holes and neutron stars with a robust mass measurement (as of March 2020). Black holes (neutron stars) discovered with gravitational waves are in blue (orange) and black holes (neutron stars) observed with electromagnetic telescopes are in magenta (green). Credit: LIGO-Virgo-KAGRA/Aaron Geller/Northwestern}
\label{f-bhgwem}
\end{figure}

Fig.~\ref{f-bhgwem} shows the stellar-mass black holes and neutron stars with a robust mass measurement (updated to March 2020 with GWTC-3). Black holes and neutron stars associated to gravitational wave events are represented, respectively, by blue and orange dots. Objects with masses in the range 3-5~$M_\odot$ (mass gap) are half-blue-half-orange dots, as we do not know exactly the value of the maximum mass for a neutron star and the minimum mass for a black hole and with current gravitational wave data we can only measure the masses of these objects, we cannot yet distinguish black holes and neutron stars from the observed waveforms. Black holes and neutron stars observed with electromagnetic telescopes are represented, respectively, in magenta and green.

%%%%%%%%%%%%%%%%%%%%%%%%%%%%%%%%%%%%%%%%%%

\subsection{Black Holes in X-ray Binary Systems}

Black hole X-ray binaries are binary systems of a stellar-mass black hole and a normal star (main sequence star or red giant). These systems are close binaries, with an orbital period normally ranging between a few hours and a few days~\cite{Casares:2013tpa}, where it is possible a significant transfer of material from the companion star to the black hole. Such a material forms an accretion disk. Thermal and non-thermal processes near the black hole produce radiation mainly in the X-ray band and we can thus see the system as an X-ray source in the sky~\cite{Bambi:2017khi}.

Black hole X-ray binaries can be grouped into two classes: {\it low-mass X-ray binaries} ($M_{\rm star} \lesssim M_\odot$) and {\it high-mass X-ray binaries} ($M_{\rm star} \gtrsim 3$~$M_\odot$), where $M_{\rm star}$ is the mass of the companion star, not that of the black hole. In the known black hole X-ray binaries, the masses of the black holes are between $\sim 5$~$M_\odot$ and $\sim 20$~$M_\odot$. 

\begin{figure}[t]
\centering
\includegraphics[width=1.0\textwidth]{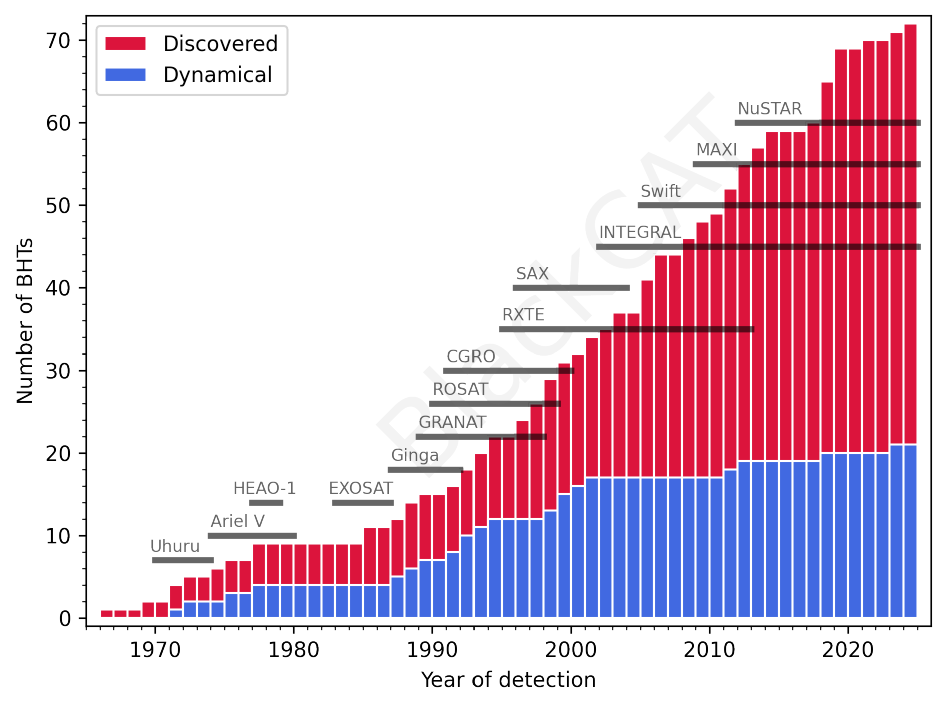}
\caption{Cumulative number of discovered Galactic stellar-mass black holes in transient X-ray sources (red bars) and cumulative number of the dynamically confirmed stellar-mass black holes (blue bars). The horizontal gray bars show the periods of activity of X-ray missions relevant for the discovery and study of black hole X-ray binaries.
Figure from the online BlackCAT catalog \url{https://www.astro.puc.cl/BlackCAT/} of Ref.~\cite{Corral-Santana:2015fud}.}
\label{f-bht}
\end{figure}

Low-mass black hole X-ray binaries are normally associated to {\it transient} X-ray sources: these systems spend most of their time in a quiescent state at a very low X-ray luminosity (which can be even too low to be detected by our X-ray observatories). The transfer of material from the companion star to the black hole normally occurs through Roche lobe overflow. This causes the formation of an accretion disk and we can see the system as an X-ray source in the sky: when this happens, we say that the system is in an {\it outburst}. An outburst normally lasts between a few weeks and a few months and then the source reenters a quiescent state for years or decades. As of now, we know about 70~low-mass black hole X-ray binaries and this number regularly increases over time because every year we may have the outburst of new systems, as shown in Fig.~\ref{f-bht}

High-mass black hole X-ray binaries are associated to {\it persistent} X-ray sources: these systems can be seen as bright X-ray sources in the sky at any time. In these systems, the companion stars are B-type or O-type stars: they have a strong stellar wind which normally permits a relatively stable transfer of material from the companion star to the black hole. As of now, we only know a few high-mass black hole X-ray binaries: Cygnus~X-1 in our Galaxy, LMC~X-1 and LMC~X-3 in the Large Magellanic Cloud, M33~X-7 in the galaxy M33, IC~10~X-1 in the galaxy IC~10\footnote{The exact nature of the compact object in IC~10~X-1 is still controversial and there is not yet unanimous consensus it is a black hole rather than a neutron star~\cite{Laycock:2015qla}.}, and NGC~300~X-1 in the galaxy NGC~300. 

\begin{figure}[t]
\begin{center}
\includegraphics[width=0.83\linewidth]{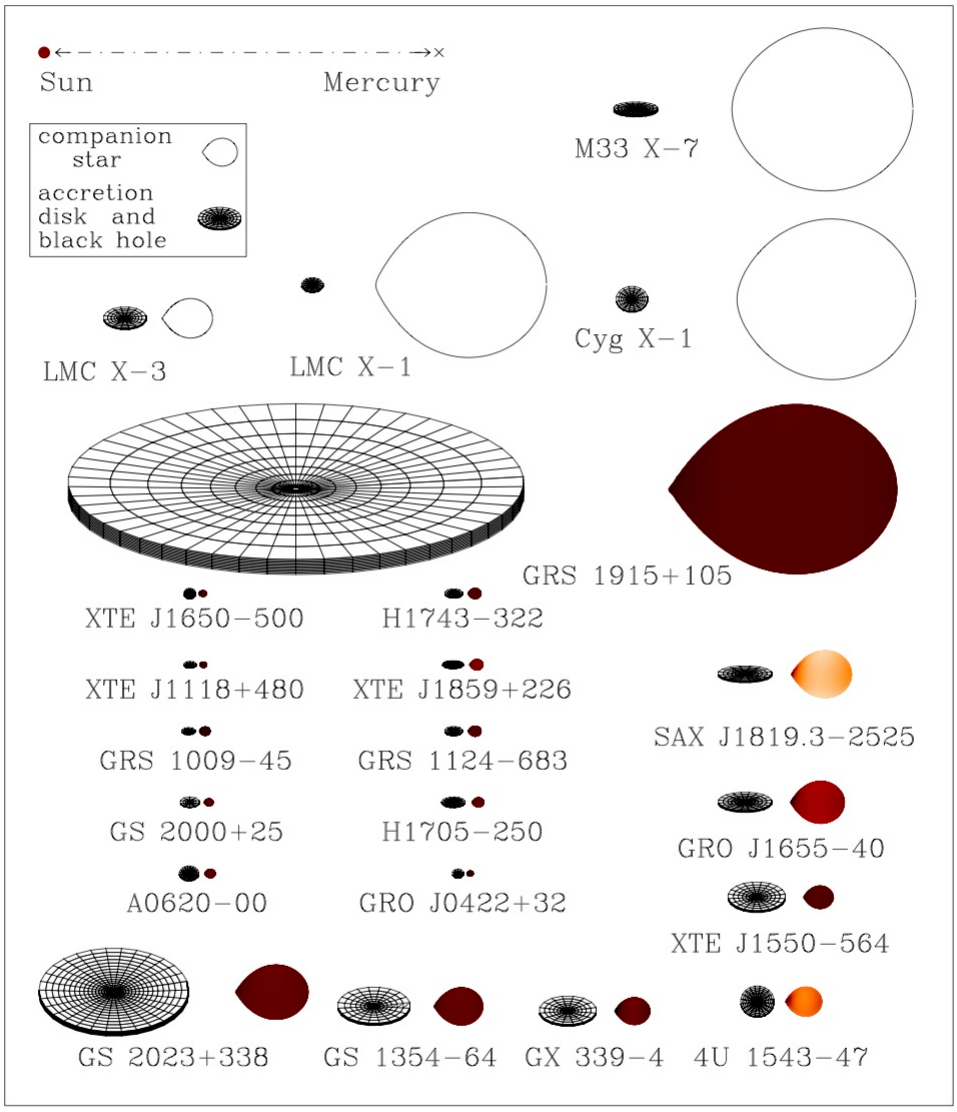}
\end{center}
\vspace{0cm}
\caption{Sketch of 22~X-ray binaries with a stellar-mass black hole confirmed by dynamical measurements. For every binary, we see the accretion disk around the black hole on the left and the companion star on the right. All accretion disks are shown at the inclination angle of the orbit of the binary. The colors of the companion stars indicate the surface temperatures of those stars (from brown to white as the temperature increases). The sizes of these binaries can be compared with the distance Sun-Mercury (about 50~millions km) reported in the top left corner of the cartoon. LMC~X-1 and LMC~X-3 are in the Large Magellanic Cloud, M33~X-7 is in the galaxy M33, and all other X-ray binaries are in our Galaxy.
Figure courtesy of Jerome Orosz. \label{f-bhb}}
\end{figure}

Peculiar systems are also possible: an example is GRS~1915+105. This is a low-mass X-ray binary and was discovered in 1992, when it entered in outburst. Since then, GRS~1915+105 has never returned to a true quiescent state and can thus be considered a persistent X-ray source. The orbital period of this binary is $\sim 34$~days, which is significantly longer than the other low-mass X-ray binaries, and this permits the black hole to have a large accretion disk, which can provide a stable mass accretion rate for a long time (see GRS~1915+105 in Fig.~\ref{f-bhb}). 

Black hole X-ray binaries can be distinguished from neutron star X-ray binaries from their spectral properties. However, for some sources the available data do not permit a clear classification of the system and there is no common agreement on whether certain objects are black holes or neutron stars. For $\sim 25$~black hole X-ray binaries, we have a dynamical measurement of the mass of the compact object; that is, we can study the orbital motion of the companion star and get that the mass of the compact object exceeds the maximum mass for a neutron stars~\cite{Casares:2013tpa}. Fig.~\ref{f-bhb} shows 22~X-ray binaries with a stellar-mass black hole confirmed by dynamical measurements.

Black holes in X-ray binaries represent only a very small fraction of the total number of stellar-mass black holes in a galaxy. In our Galaxy, models predict $10^3$-$10^4$~low-mass black hole X-ray binaries~\cite{Yungelson:2006dn,Kiel:2006hd}. High-mass black hole X-ray binaries are even more rare because the lifetime of massive stars is much shorter than that of stars with a mass similar to that of the Sun or smaller.

%%%%%%%%%%%%%%%%%%%%%%%%%%%%%%%%%%%%%%%%%%

\subsection{Black Holes in Astrometric Binary Systems}\label{ss-astrometric}

Black hole X-ray binaries are special systems, in which the black hole is close to the companion star and it is thus possible a significant transfer of matter from the companion star to the black hole and the formation of an accretion disk. Most black hole binaries are not expected to be X-ray binaries because the black hole accretion rate is too low.

Astrometric binary systems are binaries in which we only see a star and from the study of its orbital motion we can infer that such a star is in a binary system. If we can exclude the possibility that the unseen component is a normal star and infer that its mass exceeds the maximum mass for a neutron star, we can conclude that the binary system has a black hole. The past few years have seen the discovery of the following stellar-mass black holes in astrometric binary systems:

\begin{enumerate} 

\item GAIA~BH1~\cite{El-Badry:2022zih,Chakrabarti:2022eyq} (mass $M_{\rm BH} = 9.62 \pm 0.18$~$M_\odot$; orbital period of the binary $P \sim 185.6$~days; distance of the source $D \sim 480$~pc). As of now, this is the closest known black hole.

\item GAIA~BH2~\cite{Tanikawa:2022xel,El-Badry:2023pah} (mass $M_{\rm BH} = 8.9 \pm 0.3$~$M_\odot$; orbital period of the binary $P \sim 1277$~days; distance of the source $D \sim 1.16$~kpc).

\item GAIA~BH3~\cite{Gaia:2024ggk} (mass $M_{\rm BH} = 32.7 \pm 0.8$~$M_\odot$; orbital period of the binary $P \sim 11.6$~years; distance of the source $D \sim 590$~pc). As of now, this is the heaviest known stellar-mass black hole in our Galaxy and the black hole binary with the longest orbital period.

\item ALS~8814~\cite{An:2025gpa} (mass $M_{\rm BH} = \sim 15$-58~$M_\odot$; orbital period of the binary $P \sim 176.6$~days; distance of the source $D \sim 1.1$~kpc). As of now, this is the only unambiguous binary system of a Be-star and a black hole.   

\end{enumerate}

The discovery of these four black hole astrometric binaries is probably only the beginning and we can expect the discovery of many more black holes in the next years~\cite{Mashian:2017kyp,Breivik:2017cmy,Wiktorowicz:2020dvb,Janssens:2023tyt,Chawla:2023zhf}\footnote{Early studies predicted that we could discover around $10^5$ black holes in binaries with Gaia~\cite{Mashian:2017kyp}, but they neglected binary evolution. Many binaries are disrupted during or before the formation of their black hole~\cite{Breivik:2017cmy}. More recent studies predict the possibility of discovering around $10^3$~black holes in binaries with Gaia~\cite{Chawla:2023zhf}.}.

%%%%%%%%%%%%%%%%%%%%%%%%%%%%%%%%%%%%%%%%%%

\subsection{Isolated Black Holes}\label{ss-isolated}

The identification of isolated black holes is definitively more challenging. As of now, we know only a robust candidate, which is associated to the microlensing event MOA-2011-BLG-191/OGLE-2011-BLG-0462~\cite{OGLE:2022gdj}. The mass and the distance of this black hole are $M_{\rm BH} = 7.1 \pm 1.3$~$M_\odot$ and $D = 1.58 \pm 0.18$~kpc, respectively. Other potential isolated black hole candidates associated to microlensing events were reported in Refs.~\cite{Bennett:2001vh,Poindexter:2005hp,Wyrzykowski20}.

Isolated black holes can accrete from the interstellar medium and produce a detectable flux of radiation~\cite{Shvartsman,Meszaros75,McDowell85,Campana93,Fujita:1997fh,Tsuna:2018abi,Kimura:2021ayq,Murchikova:2025oio}. The mass accretion rate of an isolated black hole moving through the interstellar medium can be estimated as~\cite{Petrich89}
\begin{eqnarray}\label{eq-bondi}
\dot{M}_{\rm BH} = \lambda \frac{4 \pi M^2_{\rm BH}}{\left( v_{\rm BH}^2 + c^2_{\rm ISM}\right)^{3/2}} \mu_{\rm ISM} n_{\rm ISM} m_p \, ,
\end{eqnarray}
where $v_{\rm BH}$ is the velocity of the black hole with respect to the interstellar medium, $c_{\rm ISM}$ is the sound speed in the interstellar medium, $\mu_{\rm ISM}$ is the mean atomic mass of the interstellar medium, $n_{\rm ISM}$ is the particle number density of the interstellar medium, and $m_p$ is the proton mass. $\lambda$ is a dimensionless parameter $<1$ that reduces the ideal Bondi accretion rate because of the presence of outflows and convection (we can expect $0.01 < \lambda < 1$~\cite{Petrich89,Kaaz:2022dsa,Galishnikova:2024nsk,Kim:2024zjb}). Since $v_{\rm BH} \sim 50$~km~s$^{-1}$, the possibility of detecting isolated black holes accreting from the interstellar medium mainly depends on their distance from us, the value of $n_{\rm ISM}$, and sometimes even on the temperature of the interstellar medium through $c_{\rm ISM}$. The main challenge is to correctly identify these sources as accreting black holes. Indeed many isolated black holes may already be present in existing catalogs but have not yet been identified as such~\cite{Murchikova:2025oio}.

%%%%%%%%%%%%%%%%%%%%%%%%%%%%%%%%%%%%%%%%%%

\subsection{Black Holes in Compact Binary Systems}

\begin{figure}[t]
\centering
\includegraphics[width=1.0\textwidth]{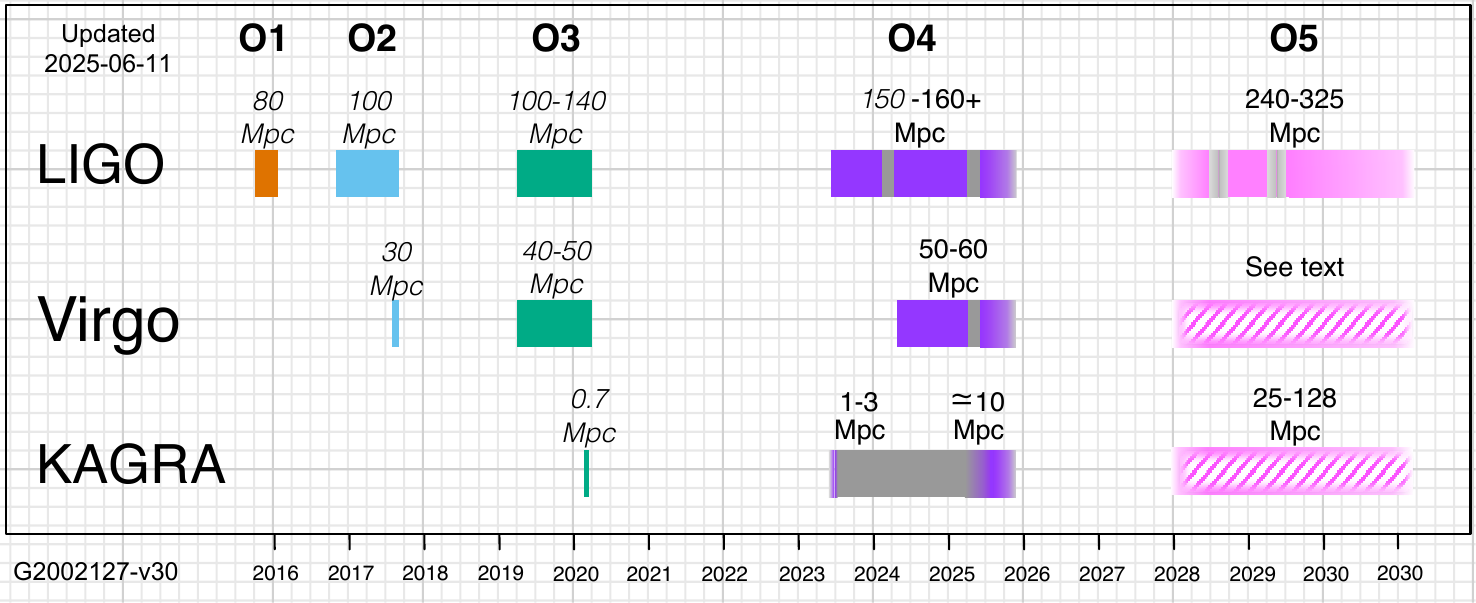}
\vspace{-0.2cm}
\caption{Timeline of the Observing Runs of the LIGO-Virgo-KAGRA Collaboration updated to June 2025. The figure also shows the binary neutron star distance for a single-detector SNR threshold of 8 in each observing run.
Figure from \url{https://observing.docs.ligo.org/plan/}.}
\label{f-O}
\end{figure}

Black holes in binary systems black hole-black hole or black hole-neutron star can be detected by current gravitational wave observatories just before the merger of the two components, when the gravitational wave signal enters the sensitivity band of our current detectors (between a few Hz and a few kHz) and is sufficiently strong. Since the first detection in September~2015~\cite{LIGOScientific:2016aoc}, about 100~events have been detected and the data are already public~\cite{KAGRA:2021vkt}. All these events were detected in the Observing Runs O1, O2, and O3 (see Fig.~\ref{f-O}). The results of the Observing Run O4 will be released soon in Fall~2025 and should include over 200~events, roughly corresponding to the detection of a new event every 3~days.

Even if we expect a large number of binary systems black hole-black hole and black hole-neutron star in our Galaxy, the lifetime of these systems is very long, so the merger of the two components is an extremely rare event. The detection of gravitational waves from the merger of two black holes (or a black hole and a neutron star) is possible only because we can monitor a large number of galaxies, and this number regularly increases with new upgrades of the detectors (as shown in Fig.~\ref{f-O}).

From the analysis of the gravitational wave signal of a specific event we can infer the properties of the system. We can measure the masses of the two objects before the merger and the mass of the final product after the merger. As of now, the quality of the data is not sufficient to reveal the internal structure of the compact objects and we can distinguish black holes and neutron stars only from the measurements of their masses. In the future, with higher-quality data, the gravitational wave signal itself should reveal if an object is a black hole or a neutron star and we should thus be able to infer the value of the Oppenheimer-Volkoff limit from observations.

%%%%%%%%%%%%%%%%%%%%%%%%%%%%%%%%%%%%%%%%%%

\section{Open Issues}

\subsection{Mass Gap}

From the measurements of masses of neutron stars and black holes in X-ray binaries, we find that compact objects with masses between $\sim 2$~$M_\odot$ and $\sim 5$~$M_\odot$ are rare (see even Fig.~\ref{f-bhgwem})~\cite{Bailyn:1997xt,Ozel:2010su,Farr:2010tu}. This would be a gap in the mass spectrum of compact objects formed from gravitational collapse, between the heaviest neutron stars and the lightest black holes. It is currently unclear if this mass gap is real or not. 

If the mass gap is real, it may be caused by the exact supernova explosion mechanism and may suggest a rapidly evolving explosion through a Rayleigh-Taylor instability~\cite{Fryer:2011cx,Belczynski:2011bn,Fryer:2022lla}. Since the mass gap is found among neutron stars and black holes in binaries, it may be also caused by the supernova natal kick and be a peculiarity of neutron stars and black holes in binaries, not of the whole neutron star and black hole population; neutron stars~\cite{Podsiadlowski:2003py} and black holes heavier than 10~$M_\odot$~\cite{Mandel:2020qwb,Burrows:2023ffl} may receive only small natal kicks on average, while low-mass black holes may receive larger natal kicks, which may disrupt their binary system~\cite{Fryer:1999ht}. On the other hand, if the gap is not real, it may be the result of, for example, selection effects~\cite{Siegel:2022gwc} or systematic errors in current mass measurements~\cite{Kreidberg:2012ud}.

The existence of such a mass gap has also been explored with gravitational wave observations~\cite{Farah:2021qom,KAGRA:2021duu}, where compact objects in the mass gap have been reported~\cite{LIGOScientific:2024elc}. From the available gravitational wave events, we see a sharp peak around 8-10~$M_\odot$ in the black hole mass spectrum but no evidence of a true mass gap at 2-5~$M_\odot$~\cite{Ray:2025aqr}. A similar conclusion may be suggested by Gaia data~\cite{Fishbach:2025bjh}, where so far we have 21~neutron star candidates, a $\sim 3.6$~$M_\odot$ black hole candidate~\cite{Song:2024tqr}, and the three black holes discussed in Subsection~\ref{ss-astrometric}: we have a first peak in the population of neutron stars at 1-2~$M_\odot$ and a second peak for $\sim 10$~$M_\odot$ black holes, but no true mass gap at 2-5~$M_\odot$~\cite{Fishbach:2025bjh}. Future gravitational wave observations and the search for black holes in Gaia's fourth data release will permit us to increase the statistics and to formulate more robust conclusions about the compact object population at 2-5~$M_\odot$.

\subsection{Black Hole Spins}

As a black hole accretes material, it changes the values of the mass and spin angular momentum. In the case of a Novikov-Thorne disk\footnote{The Novikov-Thorne model is the standard framework for geometrically thin and optically thick accretion disks~\cite{Novikov:1973kta,Page:1974he}. The model assumes that the disk is perpendicular to the black hole spin axis, the material of the disk follows geodesic circular orbits (Keplerian motion), the inner edge of the disk is at the ISCO radius, and magnetic fields are negligible. The time-averaged radial structure of the disk follows from the conservation of mass, energy, and angular momentum.}, the equilibrium black hole spin parameter is $a_*^{\rm Th} = 0.998$ ({\it Thorne limit})~\cite{Thorne:1974ve,Li:2004aq}. If the black hole spin parameter is $a_* < a_*^{\rm Th}$, the accretion process spins the black hole up. If the black hole spin parameter is $a_* > a_*^{\rm Th}$, the accretion process spins the black hole down. Magnetic fields can somewhat decreases the value of the Thorne limit~\cite{Agol:1999dn,Mummery:2025zak}. 

While there are still large uncertainties in the angular momentum transport mechanisms during the gravitational collapse of heavy stars and the formation of black holes, most models suggest that the formation of fast-rotating black holes from stellar collapses are unlikely~\cite{Woosley:2006fn,Yoon:2006fr}. Moreover, it is often thought that the value of the spin parameter of stellar-mass black holes is natal~\cite{King:1999aq,Valsecchi:2010cw,Wong:2011eg}. If the black hole is in a low-mass X-ray binary, the black hole cannot significantly change its spin parameter $a_*$ even swallowing the whole companion star, because $M_{\rm BH} \sim 10$~$M_\odot$ and $M_{\rm star} \lesssim M_\odot$. If the black hole is in a high-mass X-ray binary, the lifetime of the companion star is too short and it is impossible to transfer enough material from the companion star to the black hole even if the latter accretes at its Eddington limit. However, alternative models also exist. Fragos \& McClintock~(2015)~\cite{Fragos:2014cva} proposed that a black hole in a low-mass X-ray binary can be spun up immediately after its formation. Qin~et~al.~(2019)~\cite{Qin:2018sxk} proposed that the black hole spin in a high-mass X-ray binary may be directly related to the angular momentum of the progenitor star, which can transfer part of its envelope to the companion star and get a core with high angular momentum. 

\begin{figure}[t]
\centering
\includegraphics[width=0.75\textwidth]{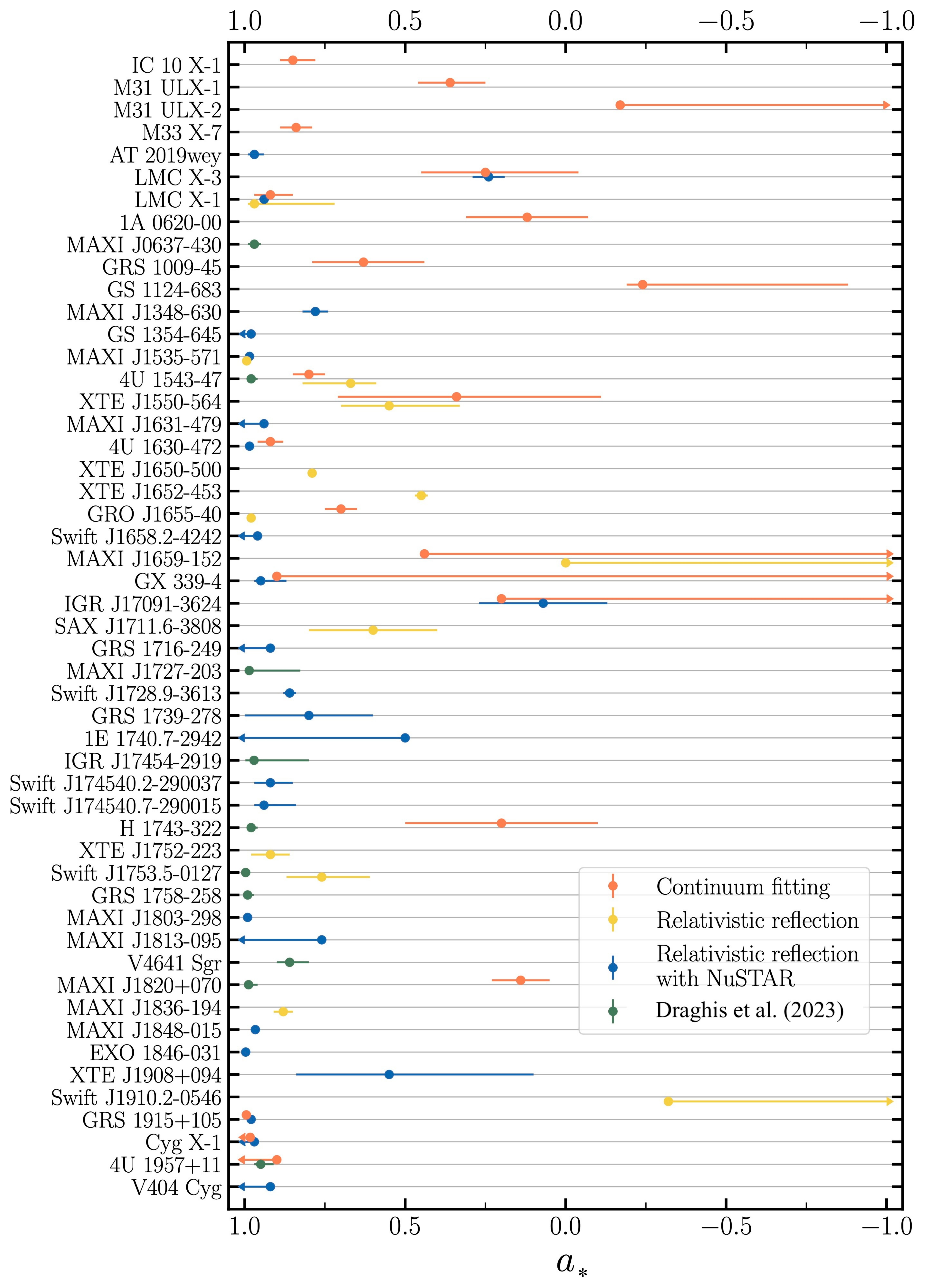}
\caption{Summary of current spin measurements of black holes in X-ray binary systems. Spin measurements inferred with the continuum-fitting method are indicated in orange, those inferred with X-ray reflection spectroscopy and without analyzing NuSTAR data are indicated in yellow, those inferred with X-ray reflection spectroscopy and analyzing NuSTAR data are indicated in blue, and the spin measurements inferred with X-ray reflection spectroscopy and analyzing NuSTAR data in Ref.~\cite{Draghis:2022ngm} are indicated in green. Figure adapted from Ref.~\cite{Draghis:2022ngm}.}
\label{f-spins}
\end{figure}

In the case of black holes in X-ray binary systems, there are two leading techniques for measuring their spins: the {\it continuum-fitting method}~\cite{Zhang:1997dy,Li:2004aq,McClintock:2013vwa} and {\it X-ray reflection spectroscopy}~\cite{Brenneman:2006hw,Dauser:2013xv,Bambi:2020jpe}. In the continuum-fitting method, we analyze the thermal spectrum of the accretion disk within the Novikov-Thorne disk model. There are five parameters in the model: the black hole mass, the black hole spin parameter, the black hole mass accretion rate, the inclination angle of the disk with respect to the line of sight of the observer, and the distance of the source. If we have independent measurements of the black hole mass, the inclination angle of the disk, and the distance of the source (for example from optical observations), we can fit the data and infer the black hole spin parameter and the mass accretion rate. X-ray reflection spectroscopy is based on the analysis of relativistically blurred reflection features produced by illumination of a cold disk by a hot corona\footnote{The {\it corona} is some hot plasma near the black hole and the inner part of the accretion disk, but its exact geometry and evolution is not yet understood well: the corona may be the hot atmosphere above the accretion disk, the material in the plunging region between the black hole and the inner edge of the disk, the base of the jet, etc.~\cite{Bambi:2020jpe}}. This technique does not directly depend on the black hole mass, mass accretion rate, and distance, and the inclination angle of the disk can be inferred from the fit. Fig.~\ref{f-spins} shows current spin measurements of black holes in X-ray binary systems. As of now, we have a spin measurement for $\sim 50$~black holes. These measurements normally rely on the assumption that the inner edge of the disk is at the ISCO radius.

Gravitational wave observations can be analyzed to estimate the posterior probability distribution of the binary parameter values, including the spins of the two initial black holes and of the final black hole~\cite{LIGOScientific:2016aoc}. Spin measurements with gravitational wave data have the advantage over X-ray measurements that are not affected by uncertainties in the astrophysical model. On the other hand, they are limited by a low signal-to-noise ratio and systematics associated with the gravitational wave modeling. 

Since X-ray measurements suggest that most black holes in X-ray binaries are fast-rotating and gravitational wave measurements found slow-rotating black holes, there is currently a debate if this is because black holes in X-ray binaries and binary black holes belong to two different object classes, there are selection effects, or one of the two methods (or both) does not provide accurate measurements~\cite{Fishbach:2021xqi,Zdziarski:2025ozs}. As shown in Fig.~\ref{f-spins}, most black holes in X-ray binary systems have a spin parameter close to 1. Draghis~et~al.~(2024)~\cite{Draghis:2023vzj} analyzed 189~NuSTAR spectra of 24~black hole X-ray binaries and found that 86\% of the sample is consistent with $a_* > 0.95$ and 100\% is consistent with $a_* > 0.7$ (1-$\sigma$ uncertainty). On the contrary, black holes detected by gravitational wave observatories have an average spin parameter $a_* \sim 0.1$ and for most black holes $a_* < 0.4$~\cite{KAGRA:2021duu}.

Low-mass X-ray binaries are not progenitors of binary black holes because the companion stars have masses $M_{\rm star} \lesssim M_\odot$ and therefore they evolve into white dwarfs, not black holes. High-mass X-ray binaries can be progenitors of binary black holes, but the observed high-mass black hole X-ray binaries formed at high metallicity: the companion stars are massive, so their lifetime is short and they have to be formed recently, which is also confirmed by the fact the masses of the black holes are $M_{\rm BH} \lesssim 20$~$M_\odot$. On the other hand, it is also true that spin measurements of black holes in X-ray binaries are affected by selection effects and systematic uncertainties. Robust spin measurements are possible for fast-rotating black holes with compact coronae, while it is difficult to distinguish fast-rotating black holes with extended coronae and slow-rotating black holes with compact coronae~\cite{Dauser:2013xv,Shashank:2025hka}.

\subsection{Is General Relativity Correct?}

Black holes are the sources of the strongest gravitational fields that can be found in the Universe today and are thus ideal laboratories for testing General Relativity in the strong field regime~\cite{Bambi:2015kza,Bambi:2024kqz}. Electromagnetic and gravitational wave observations are somehow complementary: electromagnetic tests are more suitable to probe the interactions between the matter and the gravity sectors (e.g., motion of particles in a background spacetime, presence of a fifth force, non-gravitational physics in curved spacetime, etc.) while gravitational wave tests can probe the dynamical regime (e.g., production and properties of gravitational waves, etc.).

\begin{figure}[t]
\centering
\includegraphics[width=1.0\textwidth,trim=2.2cm 1.8cm 7.5cm 17.7cm,clip]{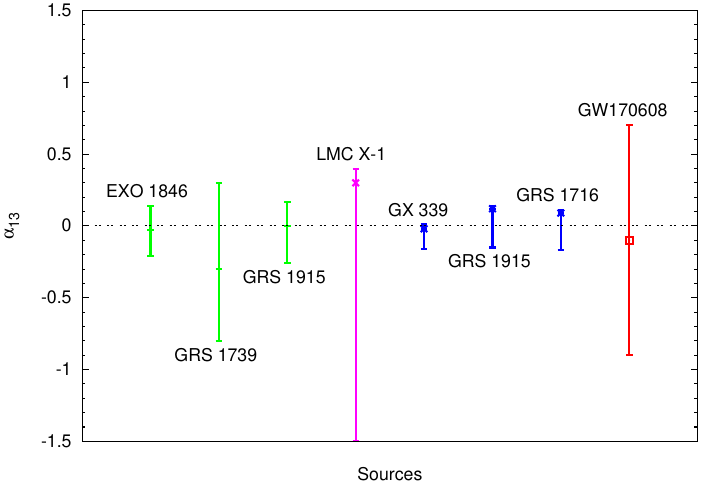}
\caption{Tests of the Kerr hypothesis -- Summary of the current 3-$\sigma$ constraints on the deformation parameter $\alpha_{13}$ from X-ray and gravitational wave observations of stellar-mass black holes ($\alpha_{13} = 0$ corresponds to the Kerr solution). The error bars in green are the best constraints on $\alpha_{13}$ from X-ray reflection spectroscopy. The error bar in magenta is inferred from the continuum-fitting method (the constraint is weak because the estimate of $\alpha_{13}$ is degenerate with the measurement of the black hole spin of the source). The error bars in blue are obtained by combining X-ray reflection spectroscopy with the continuum-fitting method for the same source. The error bar in red is the best constraint on $\alpha_{13}$ from the gravitational wave events in GWTC-3: the constraint is inferred from the inspiral phase under the assumption that the gravitational wave emission is the same as in General Relativity and only the spacetime metric can be different~\cite{Shashank:2021giy,Das:2024mjq}. The labels near the error bars refer to the names of the sources. See Ref.~\cite{Bambi:2022dtw} for more details.}
\label{f-kh}
\end{figure}

Testing the nature of astrophysical black holes is one of the most popular tests of General Relativity with black hole data and is usually referred to as the test of the {\it Kerr hypothesis}, since in General Relativity and in the absence of exotic fields we should expect that the spacetime geometry around a black hole formed from the collapse of a star should be approximated well by the Kerr solution. In the case of high-quality X-ray data, one can use X-ray reflection spectroscopy~\cite{Bambi:2016sac,Abdikamalov:2019yrr,Abdikamalov:2020oci} and/or the continuum-fitting method~\cite{Zhou:2019fcg} to test the Kerr hypothesis. The state-of-the-art of X-ray tests of the Kerr hypothesis can be found in Refs.~\cite{Tripathi:2020yts,Tripathi:2020dni,Tripathi:2021rqs,Zhang:2021ymo}; for a review on current constraints, systematic uncertainties in current measurements, and selection criteria for the sources and observations suitable for these tests, see Ref.~\cite{Bambi:2022dtw}. Fig.~\ref{f-kh} shows current constraints on the deformation parameter $\alpha_{13}$ (second Post-Newtonian corrections) from X-ray and gravitational wave data (see Ref.~\cite{Bambi:2022dtw} for more details). As of now, X-ray observations can provide stronger constraints than gravitational wave data, but it is likely that this will change with the data of the O4 cycle (see Fig.~\ref{f-O}).

With gravitational wave data, we can test specific gravity models~\cite{Yunes:2016jcc}. The gravitational wave signal does not depend on the spacetime geometry only, but even on how matter/gravity curves the spacetime, how gravitational waves are produced, and on the properties of the gravitational waves. It is not possible to test the Kerr metric in a model independent way as in the case of electromagnetic tests. Alternatively, we can look for deviations from the gravitational wave signal expected in General Relativity or we can perform a number of consistency tests~\cite{LIGOScientific:2016lio,LIGOScientific:2019fpa,LIGOScientific:2020tif,LIGOScientific:2021sio,Das:2025riq}. See Ref.~\cite{Colleoni:2024lpj} for a recent review on tests of General Relativity with gravitational wave data of the coalescence of stellar-mass black holes.

%%%%%%%%%%%%%%%%%%%%%%%%%%%%%%%%%%%%%%%%%%

\section{An Interstellar Mission to the Closest Black Hole?}

Our current theoretical models predict $\sim 10^{10}$ white dwarfs in our Galaxy (but only a small fraction of them would  be in the Galactic thin disk) and $10^8$-$10^9$~stellar-mass black holes. From observations, we know 10~white dwarfs within 25~light-years of Earth and 30~white dwarfs within 30~light-years. While the closest known black hole is GAIA~BH1 at about 1,500~light-years of Earth, there are certainly many unknown black holes closer to us.

More than 90\% of the black holes in the Galactic disk are isolated objects, without a companion star~\cite{Olejak:2019pln}. As discussed in Subsection~\ref{ss-isolated}, isolated black holes can accrete from the interstellar medium. Since the density of the interstellar medium is very low, the resulting accretion rate and accretion luminosity of the black hole are very low too. Despite that, Murchikova \& Sahu (2025)~\cite{Murchikova:2025oio} estimated that observational facilities like the Square Kilometer Array (SKA), the Atacama Large Millimiter/Submillimiter Array (ALMA), and James Webb Space Telescope (JWST) can detect isolated black holes in a warm interstellar medium within 150~light-years of Earth. However, the identification of these objects as accreting black holes is very challenging and requires multi-telescope observations. With a single telescope, we can just detect a faint source with a relatively featureless spectrum: such a source can be easily misclassified. We can identify these sources as accreting black holes only if we can observe their spectra at different wavelengths~\cite{Murchikova:2025oio}. 

If we discover a black hole within 20-25~light-years of Earth, we may send a probe to study this object and test General Relativity at a level that is likely impossible for astrophysical observations~\cite{Bambi:2025kcr}. Although very speculative and extremely challenging, such an idea is not completely unrealistic. Nanocrafts~\cite{Lubin16,Lubin22,Kuhlmey25} seem to be the most promising solution for a similar mission. A nanocraft is a gram-scale spacecraft. The main body of the probe is a gram-scale wafer with a computer processor, solar panels, navigation and communication equipment, etc. The wafer is attached to an extremely thin, meter-scale light sail, which is necessary to accelerate the probe and can be used as an antenna for communication when the nanocraft is far from Earth. Ground-based high-power lasers can accelerate the nanocraft through the radiation pressure of their laser beams. There are no specific technical problems to reach 90\% of the speed of light with this technique, but higher velocities increase significantly the total cost of the mission. If the black hole is at 20-25~light-years and the nanocraft can travel at 1/3~of the speed of light, the nanocraft can reach the black hole in 60-75 years, performs some experiments in the strong gravitational field of the black hole, and we need 20-25 more years to receive the results of the experiments~\cite{Bambi:2025kcr}. While this is currently only a rough idea that requires more studies, if there is really a black hole within 20-25~light-years of Earth it may just be an issue of time to reach the necessary technology to send a probe and study the object.

%%%%%%%%%%%%%%%%%%%%%%%%%%%%%%%%%%%%%%%%%%

\section{Concluding Remarks}

Stellar-mass black holes are the natural evolution of heavy stars. In our Galaxy, we expect $10^8$-$10^9$~stellar-mass black holes formed from stellar collapse, but currently we know less than 100~objects. Their detection is indeed extremely challenging and eventually possible only in very special conditions. 

The past 10~years have significantly changed our understanding of the physics and astrophysics of stellar-mass black holes. Since~2015, we can study these objects with gravitational waves and current gravitational wave detectors can roughly observe a new coalescence of black holes every 3~days. In the past few years, we have started discovering non-interacting black holes, like black holes in astrometric binaries or isolated black holes such as MOA-2011-BLG-191/OGLE-2011-BLG-0462. If we go back to 10~years ago, we only knew stellar-mass black holes in X-ray binaries and we claimed they were black holes simply because they were dark and compact objects exceeding the maximum mass for neutron stars; there was no evidence that these objects were the Kerr black holes predicted by General Relativity. Today we test the nature of stellar-mass black holes with X-ray and gravitational wave observations: current constraints may not be very stringent, but they can be improved with future observations. 

The purpose of this review article was to provide a short and updated overview on the physics and astrophysics of stellar-mass black holes. It does not pretend to be complete and provides a list of recent and less-recent references for every topic. Considering the quick evolution of this research field, parts of this review article will likely become out of date soon.

%%%%%%%%%%%%%%%%%%%%%%%%%%%%%%%%%%%%%%%%%%

\vspace{6pt}

%%%%%%%%%%%%%%%%%%%%%%%%%%%%%%%%%%%%%%%%%%

\acknowledgments{The author thanks Jorge Casares and Jesús M. Corral-Santana for Fig.~\ref{f-bht} and Jerome Orosz for Fig.~\ref{f-bhb}.}

\funding{This work was supported by the National Natural Science Foundation of China (NSFC), Grant No.~12261131497 and No.~12250610185.}

\dataavailability{No new data were created in this work.}

\conflictsofinterest{The author declares no conflicts of interest.}

%%%%%%%%%%%%%%%%%%%%%%%%%%%%%%%%%%%%%%%%%%
%\isPreprints{}{% This command is only used for ``preprints''.
\begin{adjustwidth}{-\extralength}{0cm}

\reftitle{References}

%%%%%%%%%%%%%%%%%%%%%%%%%%%%%%%%%%%%%%%%%%
\PublishersNote{}
%\isPreprints{}{% This command is only used for ``preprints''.
\end{adjustwidth}
\end{document}